\documentstyle[12pt]{article}
\begin{document}
\begin{center}{\bf Leading/nonleading Charm Hadroproduction in the
Quark-Gluon String Model} \end{center}
\begin{center}
O.I.Piskounova
\end{center}
\begin{center}
{\it Joint Institute for Nuclear
Research, Dubna 141980, Russia}
\end{center}
\vskip 1em

\begin{abstract}
Leading charm production in reaction $\pi^- N \rightarrow DX$
is discussed.Three experiments at Fermilab and CERN have measured a rather
large enhancement in production rate of leading $D^-$-mesons over nonleading
$D^+$.A good description of these results is obtained in Quark Gluon
String Model (QGSM) if the idea of 'intrinsic charm' is involved.
The weight of the $c\bar{c}$ pairs in the quark sea of colliding hadrons is
estimated from the experimental data. The comparison with other theoretical
models is also carried out.

\end{abstract}

\section{Introduction}

Leading effects in charm hadroproduction have been discussed \cite{rew1,
rew2} for many years of research in this field. Measurements have been made
recently in reactions like $\pi^- A \rightarrow DX$ for the momenta of
incident $\pi^{-}-$beams from 250 GeV/c to 500 GeV/c \cite{wa82,e769,e791}.
An essential enchancement of leading $D^-$- over nonleading $D^+$-meson rates
has been observed, especially in the region of Feynman's x$_f\rightarrow$1.
This asymmetry in the spectra of charmed mesons is defined everywhere as
\begin{equation}
A(x)=\frac{dN^{D^-}/dx-dN^{D^+}/dx}{dN^{D^-}/dx+dN^{D^+}/dx}.
\end{equation}
It indicates a strong dependence on the valence quark composition
of the beam particle, the so called "beam dragging" effect. Charmed mesons,
$D^-$(d$\bar{c}$) and $D^0$($\bar{u}$c) in this case produced
by a recombination of the $\bar{u}$ or d valence quarks from $\pi^-$ with the
$c\bar{c}$ pair from fission of a string,have harder spectra than for
the other D-mesons.

The magnitudes of A(x) measured in WA82\cite{wa82}, E769\cite{e769} and E791
\cite{e791} experiments in the narrow range of $\sqrt{s}$=23-32 GeV have
close values.

The models previously applied to describe the asymmetry dependence, were not
very successful, see ref.\cite{bednjakov}, for instance. The first order QCD
calculation could not produce any asymmetry because it did not take into
account the quark content of the beam or target particles. The model \cite
{pythia} based on Lund-type string fragmentation and implemented as the
Monte-Carlo programme PYTHIA usually overestimates the leading effect. There
is a clear reason for such a behavior: strong asymmetry is obtained because
only valence quarks can be at the ends of the strings. The fragmentation of
strings gives a strong dependence between the quark contents of the produced
leading and projectile particles.

S.J.Brodsky et.al.\cite{brodsky} was the first to have put forward the idea
that 'intrinsic charm' suppresses the leading/nonleading asymmetry because the
c and $\bar{c}$ quarks in the projectile particle produce both the leading and
nonleading D-mesons.

The Quark Gluon String Model \cite{prev papers} to  be discussed
in this note has described well the most of experimental data on
the x$_{f}$-spectra of charmed particles \cite{charm}. Intrinsic charm
can be taken into account in the framework of this model as an admixture of
the $c\bar{c}$ pairs in the quark sea of interacting particles. This paper
considers the effects to be expected if some fraction of the charmed sea quarks
is
involved. It should be mentioned that the nuclear target will not be taken
into account. Only pion-proton collisions will be considered.

\section{Intrinsic charm in QGSM}

Spectra of charmed mesons produced in pion-proton reactions are determined
(in terms of QGSM) by the sum of chain distributions over all the cut $\it
n$-pomeron diagrams as shown in fig.1. The D-meson distribution in each
diagram consists of the contributions from the valence quark-antiquark or
antiquark-diquark chains and $\it{2(n-1)}$ sea quark-antiquark chains:
\begin{eqnarray}
\varphi _{n}^{D}(s,x)=a_{0}^{D}(F_{q}^{(n)}(x_+) F_{qq}^{(n)}(x_{-})+
F_{\bar{q}}^{(n)}(x_{+}) F_{q}^{(n)}(x_{-})\\ \nonumber
+2(n-1)F_{q_{sea}}^{(n)}(x_{+})F_{\bar{q}_{sea}}^{(n)}(x_{-})),
\end{eqnarray}
where $a_{0}^{D}$ is the density of the D-meson formation in the center
of the chain absolutely independent on the kind of D's.

The leading/nonleading difference is expressed mostly in the form of
fragmentation functions convoluted with the structure functions
of quarks at the ends of chains in formulae for F$_{q,\bar{q},q_{sea}}$, as
in the following equation:
\begin{equation}
F_{i}(x_{\pm})=\int_{x_{\pm}}^{1} f_{i}(x_{1})\frac{x_{\pm}}{x_{1}}
{\cal D}_{i}^{D}(\frac{x_{\pm}}{x_{1}})dx_{1}.
\end{equation}

The nonleading fragmentation function $\cal{D}(\it{z})$ is written in the
usual QGSM form :
\begin{equation}
{\cal D}_{d}^{D^{+}}(z)=\frac{1}{z}(1-z)^{-\alpha_{\psi}(0)+\lambda+
2(1-\alpha_{R}(0))},
\end{equation}
where  $\lambda$=2$\alpha_{D^{*}}^{\prime}$(0)$\bar{p_{\bot D^{*}}^{2}}$,
$\alpha_{\psi}(0)=-2.$  \cite{charm}.

The leading type fragmentation function contains the factors
important for asymmetry:

\begin{equation}
{\cal D}_{d}^{D^{-}}(z)=\frac{1}{z}(1-z)^{-\alpha_{\psi}(0)+\lambda}
(1+a_{1}^{D}z^{2}).
\end{equation}

The additional 2(1-$\alpha_R(0)$) in (4) means that at least one pair of
ordinary quarks must be produced in addition to the valence quark at the
top of the chain to obtain a nonleading D-meson ($\alpha_R(0)$=0.5 is the
intercept of an ordinary quark Regge trajectory). The $a_1^Dz^2$ term is
introduced in \cite {kaidalov} to provide a transition between
probabilities of the D$^{-}$ production at z$\rightarrow$0 and z$\rightarrow$1.

Such a difference between "favoured" and "unfavoured" fragmentations leads to
a large asymmetry increasing with x.This asymmetry will be suppressed with the
presence of the $c\bar{c}$ quark pairs in the quark sea of pion.
These quarks have the following distributions:
\begin{equation}
f_{c(\bar{c})}^{(n)}=C_{c_{sea}}^{(n)}\delta_{c(\bar{c})}x_1^{-\alpha_{\psi}(0)}
(1-x_1)^{\alpha_R(0)-2\alpha_N(0)+(\alpha_R(0)-\alpha_{\psi}(0))+n-1}
\end{equation}
where $x_1$ is the momentum fraction of the $c(\bar{c})$ quarks and
$\delta_{c(\bar{c})}$ is the weight of the charmed quark pairs in the quark
sea.

The fragmentation function for the chain attached to the $c(\bar{c})$ sea quark
is of leading type, for example:

\begin{equation}
{\cal D}_{c(\bar{c})}^{D^{-}(D^{+})}(z)=\frac{a^D_f}{a^D_0z}z^3(1-z)^{-\alpha_
R(0)+\lambda}.
\end{equation}
where $a^D_f$ is of order 1.

This phenomena will cause a fall in asymmetry in the region of $\it{x}$ where
the charmed quark sea pairs are distributed.As it is evident from eq.(6),charm
quarks are not spread out in narrow region of $x_1$,then the suppression of
asymmetry will be observed more or less in the whole range of x.

\section{Comparison between models and data}

The asymmetry predicted in QGSM is shown as a solid line in figs.2 and
3. The value of the $c\bar{c}$ admixture is $a^D_f\delta_{c(\bar{c})}$=0.1.
The behavior of each model is well illustrated according to description given
above. The PYTHIA calculations give everywhere the highest asymmetry curves
because the suppression caused by intrinsic charm is not taken into account.
QGSM curves are of the similar form , but the agreement with data is better in
the absolute values.There is a fixed fraction of intrinsic charm and specific
distribution in $\it{x}$ in the model of ref.\cite{brodsky}.That is why the
curve of the model in fig.3 is considerably lower than the experimental data
in region $\it{x}$ near 0.2. The better fits are possible as in PYTHIA and
the model of ref.\cite{brodsky}.

\section{Conclusions}

Thus, as it was shown in paragraph 3 a small fraction of intrinsic charm can
lead to a satisfactory description of experimental results. Such fraction was
included into QGSM as an additional phenomenological parameter and has been
estimated from the comparison with experimental data as $\delta_{c(\bar{c})}$
=0.1. This picture can be improved if to take into account the influence
of nuclear target effects. To obtain an important information on properties of
hadronic wave function at high energies, it is necessary to provide a more
precise evaluation of the intrinsic charm structure function.

\newpage

\begin{picture}(140,140)

\end{picture}
\begin{center}{\it Fig.1. Diagram with n-pomeron exchange.}\end{center}

\begin{picture}(140,140)

\end{picture}
\begin{center}{\it Fig.2. WA89 and E769 results....................Fig.3. E791
data.}
\end{center}

\end{document}